\documentclass[%
 preprint, %linenumbers,
nofootinbib,
 amsmath,amssymb,
 aps, physrev,
]{revtex4-2}

\usepackage[english]{babel}

\renewcommand{\selectlanguage}[1]{}

\usepackage{graphicx}% Include figure files
\usepackage{dcolumn}% Align table columns on decimal point
\usepackage{bm}% bold math
\usepackage{hyperref}% add hypertext capabilities
\usepackage[mathlines]{lineno}% Enable numbering of text and display math
\begin{document}

\preprint{APS/123-QED}

\title{\textbf{On the Numerical Stability of the Diffusion Coefficient in Microscopic Simulations} 
}% 

\author{Yurovsky V.O.}
  \affiliation{Faculty of Physics, M. V. Lomonosov Moscow State University (MSU), 119991 Moscow, Russia}%Lines break automatically or can be forced with \\
  \altaffiliation[Also at ]{Skobeltsyn Institute of Nuclear Physics, M. V. Lomonosov Moscow State University (MSU), 119991 Moscow, Russia}
  \email{Contact author: yrovskyvladimir@gmail.com}
\author{Kudryashov I.A.}%
  \affiliation{Skobeltsyn Institute of Nuclear Physics, M. V. Lomonosov Moscow State University (MSU), 119991 Moscow, Russia}%


\date{\today}% It is always \today, today,

\begin{abstract}
Nowadays, the calculation of the Galactic Cosmic Rays diffusion coefficient with direct microscopic numerical simulations is a widespread approach. In this work, we investigated the numerical limits for such calculations and demonstrated that modern computations are affected by the influence of numerical errors. We found that velocity errors have a greater impact on the result than spatial ones.
%\begin{description}
%\item[Keywords]
%Galactic Cosmic Rays, Diffusion, Numerical Simulations
%\end{description}
\end{abstract}

\keywords{Galactic Cosmic Rays, Diffusion, Numerical Simulations}%Use showkeys class option if keyword
\maketitle

%\tableofcontents

\section{Introduction}
\label{sec:intro}

Galactic cosmic rays (GCRs) are energetic charged particles that pervade the interstellar medium (ISM) and serve as unique probes of both distant astrophysical accelerators and the magnetized, turbulent plasma through which they propagate. The transport of GCRs from their sources to Earth is a main process that shapes their observed energy spectra, composition, and arrival directions. Observations have established the following phenomenological picture: the extreme isotropy of the GCR flux \cite{AMS}, the overabundance of light secondary nuclei such as boron and beryllium relative to their primary nuclei \cite{Dampe}, and the inferred multi-million-year residence time within the Galaxy \cite{GCRLifetime}. These phenomena indicate that the particles do not stream freely but instead execute a highly irregular, diffusive journey. These phenomena can be naturally explained if the motion of GCRs is dominated by resonant pitch-angle\footnote{The Pitch-angle description is not applicable for sufficiently large energies, when the gyroradius of the particle exceeds the characteristic size of the magnetic field inhomogeneities, i.e., the correlation length of the magnetic field. For larger energies, another description is preferable. For a discussion of cosmic ray transport regimes see, for example, \cite{shalchi2009nonlinear}.} scattering off magnetohydrodynamic (MHD) turbulence which reduces the complex single-particle trajectory to a classical random walk.

The standard theoretical framework for describing this transport is a Fokker–Planck equation for the nearly isotropic part of the GCR distribution function. Under the diffusion approximation, which is valid when the scattering frequency is large compared with macroscopic gradient scales, the spatial part of the equation reduces to a second-order parabolic term featuring a diffusion tensor $D_{ij}$. This diffusion tensor is conventionally parameterized by its components parallel and perpendicular to the mean magnetic field, and its energy dependence is inferred from secondary-to-primary ratios such as B/C \cite{Maurin_2001}. The best-known numerical model within the framework of normal diffusion is currenlty the GALPROP \cite{GALPROP} numerical code which however implements only isotropic diffusion. Anisotropy of the GCR diffusion is addressed by DRAGON \cite{Dragon} code or in works \cite{OurWork} which take into account the mid-range realistic structure of the galactic magnetic fields. While this macroscopic diffusion model has been remarkably successful in fitting large-scale GCR data, it is fundamentally phenomenological: Diffusion coefficients are not predicted from first principles but instead are tuned to match observations, with guidance from simplified theories such as quasilinear theory \cite{Jokipii1966-yb}. Some works \cite{Kuhlen2025-fg} try to overcome this issue and obtain transport parameters ab initio from results of the direct numerical simultaions. Perhaps, CRPROPA \cite{CRPROPA} is the most famous numerical code which implements microscopic numerical simulations of the GCR transport.

However, a fundamental question has received surprisingly little systematic scrutiny: is the diffusion observed in these numerical experiments a true physical consequence of the prescribed magnetic fields, or could it be, at least partially, a numerical artifact? This is not merely a technical curiosity. Test-particle simulations involve a chain of numerical choices -- discrete time integration schemes, interpolation of fields onto particle positions, finite spatial resolution of turbulence spectra, artificial boundary conditions, and limited integration times -- each of which can introduce spurious randomness, artificial trapping, or truncation of long free flights. Unless carefully controlled, such artifacts can lead to incorrectly “measured” diffusion coefficients, which are then included in large-scale phenomenological models, potentially distorting our understanding of GCR propagation and its comparison with observational data.

In this paper, we address this issue head-on. We employ ab initio simulations of the GCR microscopic transport and study the stability of transport parameters with respect to the technical parameters of the integration algorithm. Also, we introduce some artificial errors into the integration algorithm to find the most sensitive parts of the system.

The paper is organized as follows. Section \ref{sec:numericalSetup} describes our simulation setup. In Section \ref{sec:solutionStability} and Section \ref{sec:dStability} we study stability of the trajectories and the behavior of the so-called running diffusion coefficient when varying the time step. Sections \ref{sec:spatialError}, \ref{sec:velocityError} examine the influence of spatial and velocity errors by introducing an additional artificial error into the simulation. The estimation of algorithm-inherent errors is presented in Section \ref{sec:errorEstimation}. Finally, Section \ref{sec:conclusion} presents a discussion of the results and conclusion.

\section{Numerical setup}
\label{sec:numericalSetup}

\subsection{Magnetic field}

Modern models of the Galactic magnetic field, like UF23 \cite{UF23_GMF}, distinguish several components of the field. In this work a simple model, based on \cite{magneticField}, is used. Magnetic field is modeled as a sum of turbulent and regular components, denoted as $\mathbf{b}(\mathbf{r})$ and $\mathbf{B}_0$, respectively. The direction of $\mathbf{B}_0$ is arbitrarily selected along z axis.

\begin{equation}
    \mathbf{B}(\mathbf{r}) = \mathbf{B}_0 + \mathbf{b}(\mathbf{r})
\end{equation}

Turbulent part of the magnetic field $\mathbf{b}(\mathbf{r})$ is modeled as a gaussian random field with random phases. Magnetic field is represented as a finite sum of modes with randomly generated uniformly distributed phases and wave vector directions. Amplitudes of modes obey the Kolmogorov spectrum law. 

\begin{equation}
    \mathbf{b}(\mathbf{r}) = \sum_NA_n\mathbf{P}_n\cos(\mathbf{k}_n\mathbf{r} + \phi_n)
\end{equation}

The number of modes used in calculations in this work equals $500$. The minimum wave vector corresponds to the wavelength of $100$ AU, the maximum wave vector corresponds to the wavelength of $100$ pc. The RMS value of the turbulent component is equal to the amplitude of the regular component. This common value is chosen as $6$ $\mu$G, which is in agreement with modern models of the Galactic magnetic field. 

Exact values of the magnetic field parameters are not important for achieving the article's goals. What is important is that we need to study transport in a regime where particles are pinned to the magnetic field lines and the selected parameter values and particle energies, used in simulations, satisfy this requirement.

\subsection{Simulation setup}

We numerically integrate the following equations for particle trajectories

\begin{equation}
\begin{cases}
\frac {d \boldsymbol r}{dt} = c \tilde{\boldsymbol v} \\
\frac{ d \tilde{\boldsymbol v}}{dt} = \frac{qc^2}{E} [\tilde{\boldsymbol v} \times \boldsymbol B],
\end{cases}
\end{equation}

These equations are a rewritten form of Newton-Lorentz equations with energy explicitly shown. For getting solution we are using the Cash-Karp method. Despite that a common recommendation is to select Boris method (See for example \cite{BPGOOD} or appendix in \cite{Kuhlen2025-fg}), Cash-Karp method is the fourth order method while the Boris-Push method has just second order accuracy and as we will see later, even fourth-order accuracy is not enough for good trajectory stability.

In each run we simulate some number of particles (most commonly $64$, but it will be mentioned explicilty for each result) started from one point with evenly distributed velocity directions.

The main step size in AU is selected to be equal to the particle energy in TeV. For example, for a particle with energy equal to $100$ TeV, the step size is $100$ AU. For the selected magnetic field amplitude, such choice of the step size is approximately 1/30 of the gyroradius for any energy and still allows sufficiently long computations, and thus is a good compromise for the main steps. We will be studying other step sizes and will usually refer to them as "relative" step sizes relatively to the main step size. The selected main step size will be referred to as a unit step size.

\subsection{Diffusion coefficients}

In this work we discuss the running diffusion coefficient which is defined as follows

\begin{equation}
    D_{ij}(t)=\frac{<x_i(t)x_j(t)>}{t} ,
\end{equation}
where D$_{ij}$ is the component of the diffusion tensor (which arises due to anisotropic magnetic field), x$_i$ is the component of the radius vector of the particle, t is time, and angular brackets $<>$ denote ensemble averaging.

\subsection{Statistical processing}

All results are obtained using ensemble averaging. Confidence intervals are obtained using bootstrap method, number of bootstrap samples is $1000$. The shown interval is between $2.5$ and $97.5$ percentiles and thus represents $2 \sigma$ confidence interval.

\section{Numerical stability of the solution}
\label{sec:solutionStability}

\begin{figure}
    \centering
    \includegraphics[width=1\linewidth]{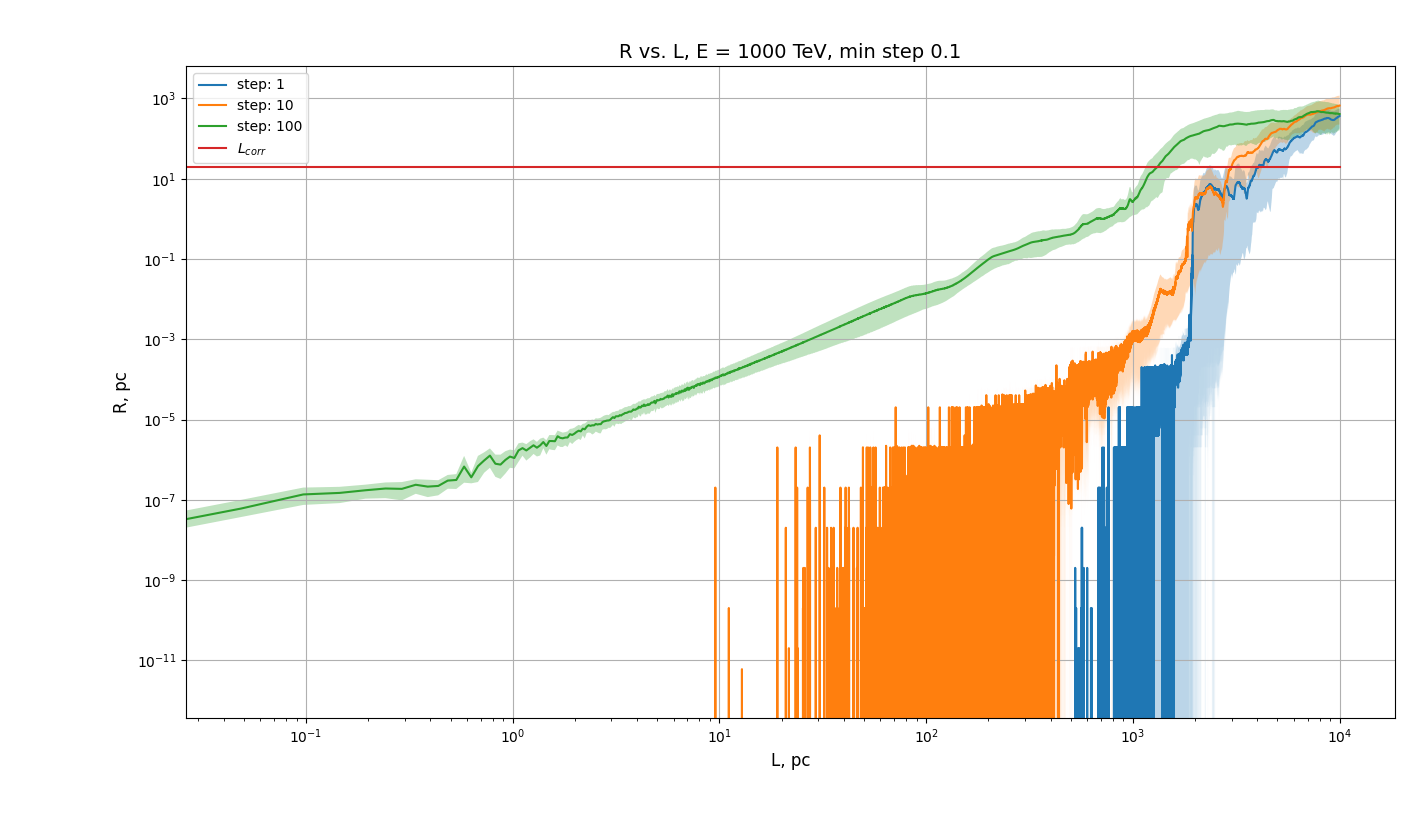}
    \caption{RMS displacement vs path length dependency. The red line denotes the correlation length of turbulent part of the magnetic field.}
    \label{fig:RL_1e4pc}
\end{figure}

To study the numerical stability of the algorithm we simulated trajectories of particles with different step sizes. The initial states and the magnetic field remained the same. To represent the trajectory divergence we plotted the RMS absolute difference of the particle trajectory obtained with a certain step size and the same trajectory obtained with the minimal used step size:

\begin{equation}
    R(l) = <(\mathbf{r}_{i}(l)-\mathbf{r}_{0}(l))^2>  ,
\end{equation}
where $\mathbf{r}_i(l)$ is the radius vector of a particle obtained with step $i$ at path length\footnote{Because the selected energies are much larger than the proton's rest energy, particle's velocity is approximately equal to the speed of light, and thus we can determine path length as $l = c \cdot t$, where c is the speed of light and t is the simulation time} l and $\mathbf{r}_0(l)$ is the radius vector of the same particle obtained with the minimal step size at the same path length. In these calculations we used relative\footnote{Recall that a relative step size of 1 corresponds to a step size equal to 1/30 of the gyroradius.} step sizes $0.1, 1, 10, 100$, the path length was $10$ kpc, and particle energies were $1000$ TeV, 16 particle trajectories were calculated. Results are presented in Figure \ref{fig:RL_1e4pc}. Here we can see that for selected conditions RMS distances between particle trajectories exceed one correlation length of the magnetic field at path lengths on the order of $1$ kpc, and these trajectory divergences for step sizes $10$ and $1$ do not show significant difference after that path length. This means that even with smaller step sizes the trajectory convergence is hardly achievable.

\section{Numerical stability of the diffusion coefficient}
\label{sec:dStability}

\begin{figure}
    \centering
    \includegraphics[width=1\linewidth]{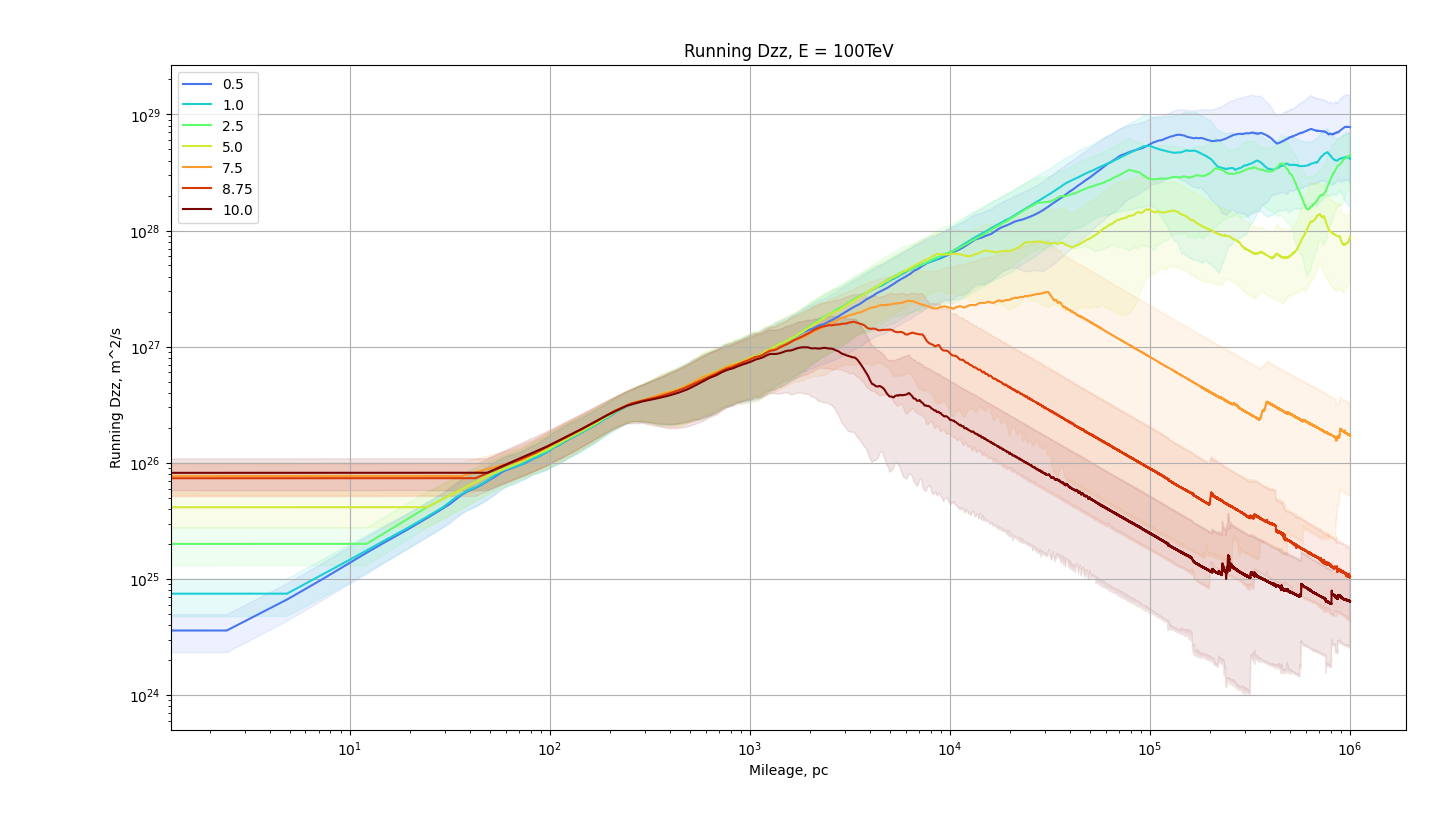}
    \caption{Running Dzz vs path length dependency for different step sizes.}
    \label{fig:DzzStepErr}
\end{figure}

The necessity of the trajectory convergence for correctness of the diffusion coefficient calculations is not proven and is not obvious due to the stochastic nature of diffusion. In other words, it is still possible that we can obtain reliable statistical parameters from non-converging trajectories, so we studied the behavior of the running diffusion coefficient for different step sizes. In these calculations we used relative step sizes $0.5, 1, 2.5, 5, 7.5, 8.75, 10$, the path length was $1000$ kpc, and particle energies were $100$ TeV, and 16 particle trajectories with random initial velocity directions were calculated. For each step size the same set of initial velocities was used. The most interesting diffusion tensor component is the diffusion coefficient along z axis, because diffusion in this direction is strongly affected by the presence of the regular magnetic field and calculated diffusion coefficients in this direction are several orders of magnitude higher than in perpendicular direction. Results are presented in Figure \ref{fig:DzzStepErr}. Here we can see that our selected main step size is close to the critical step size, at which the behavior of the running diffusion coefficient changes dramatically. The divergence of the curves for small path lengths is due to the absence of trajectory points\footnote{For larger step sizes we have no intermediate points, while for smaller step sizes we do have intermediate points.} and the odd transformation of the curves in double logarithmic scales\footnote{Initial points are the same and are located at (0, 0), which is transformed into (-$\infty$, -$\infty$) on a double logarithmic scale.}. Also, we can notice that the smaller the step size, the earlier the approach of diffusive behavior\footnote{This can be determined by the constant behavior of the curve.} and the smaller the asymptotic diffusion coefficient. It is obvious that numerical errors affect the obtained result. In the following sections we are going to reveal which error type has the largest impact.

\section{Spatial error impact}
\label{sec:spatialError}

\begin{figure}
    \centering
    \includegraphics[width=1\linewidth]{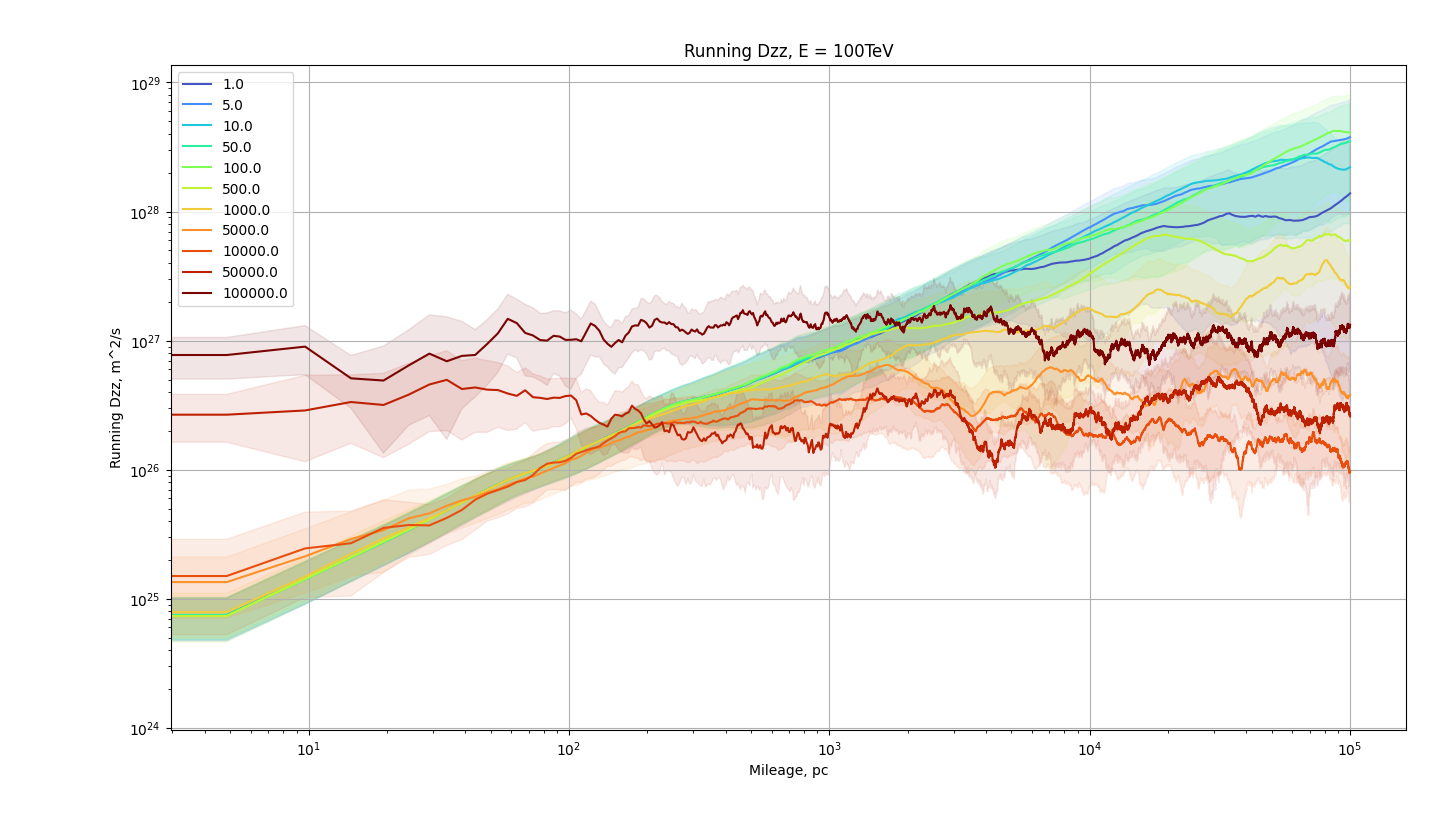}
    \caption{Running Dzz vs path length dependency for different spatial errors. Different colors denote different artificial error magnitudes, labels are in AU.}
    \label{fig:DzzRErr}
\end{figure}

To study the influence of a spatial error on running diffusion coefficient calculations we develop the following approach. After each step of the integration algorithm we add random displacement to the particle position in a uniformly distributed direction with the constant selected magnitude. At certain magnitudes this artificial error should lead to the diffusion guided by the random displacement itself. In these calculations we used a relative step size $1$, magnitudes of random displacement $1, 5, 10, 50, 100, 500, 1000, 5000, 10000, 50000, 100000$ AU, the path length was $100$ kpc, and particle energies were $100$ TeV, 16 particle trajectories were calculated. Results are presented in Figure \ref{fig:DzzRErr}. Here we can see that the minimal displacement amplitude at which significant changes become observable is between $100$ and $500$ AU which is, at least, not smaller than the step size itself ($100$ AU). So, the influence of spatial error, at least in the form introduced above, is nearly negligible. Also, we can notice that at some absurdly large artificial error magnitudes particle diffusion is guided purely by introduced artificial error.

\section{Velocity error impact}
\label{sec:velocityError}

\begin{figure}
    \centering
    \includegraphics[width=1\linewidth]{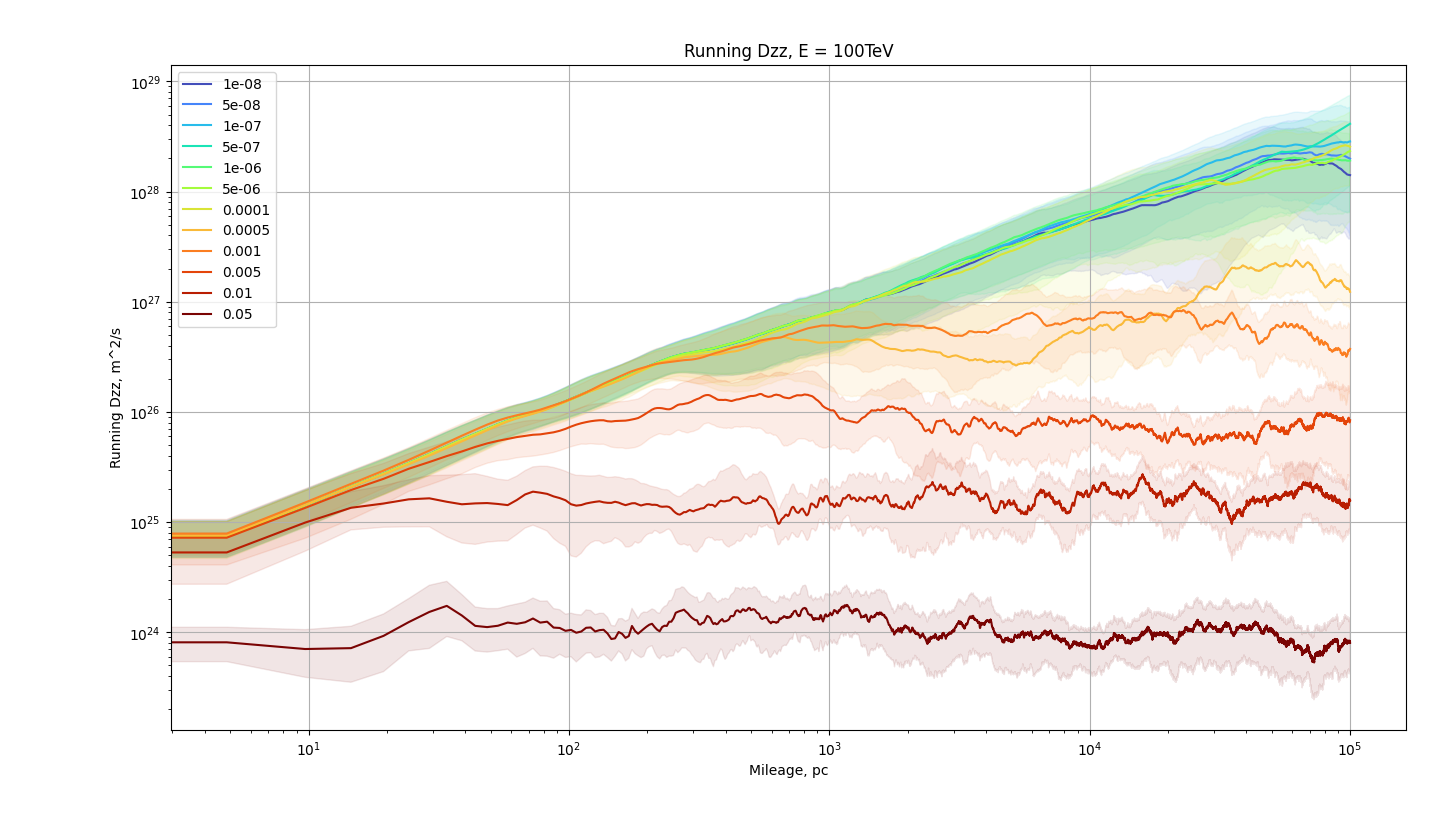}
    \caption{Running Dzz vs path length dependency for different velocity errors. Different colors denote different random rotation angles, labels are in radians.}
    \label{fig:DzzVErr}
\end{figure}

To study the velocity error influence we developed another approach. After each step of the integration algorithm we rotated the velocity of the particle in a random direction on a selected constant angle. In these calculations we used a relative step size $1$, angles of random rotation from $1 \cdot 10^{-8}$ up to $0.05$ radian, the path length was $100$ kpc, and particle energies were $100$ TeV, 16 particle trajectories were calculated. The results are presented in Figure \ref{fig:DzzVErr}. Here we can see that at some magnitude (between $5 \cdot 10^{-6}$ and $5 \cdot 10^{-4}$ radians) the added error significantly influences the transport. We can observe that the more velocity error the less diffusion coefficient is. It can be easily explained with pitch-angle decorrelation. So, velocity error is the main error source in the simulations. 

\begin{figure}
    \centering
    \includegraphics[width=1\linewidth]{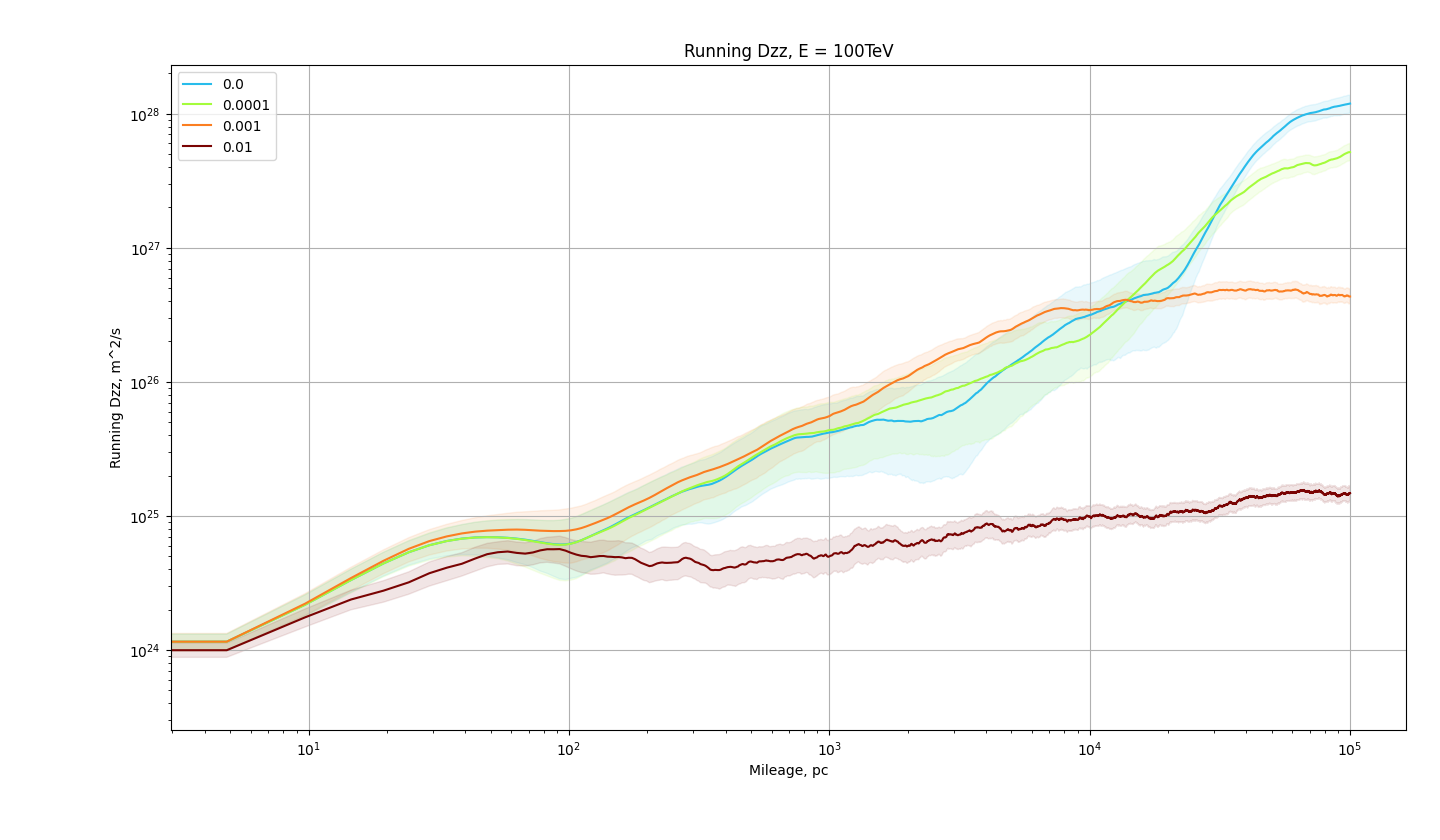}
    \caption{Running Dzz vs path length dependency for different velocity errors. Different colors denote different random rotation angles, labels are in radians.}
    \label{fig:DzzVErrManyPts}
\end{figure}

Also we studied velocity error influence with bigger statistics but with less number of different error angles. In this calculation the number of particle was $448$, other parameters were the same. Results are presented in Figure \ref{fig:DzzVErrManyPts}. Taking into account wide confidence intervals, the behavior of the running diffusion coefficients remains the same. %At first glance this result may look strange, but if we take into account wide confidence interval then our conclusions about velocity error influence remain unchanged.

\section{One step error estimation}
\label{sec:errorEstimation}

\begin{figure}
    \centering
    \includegraphics[width=1\linewidth]{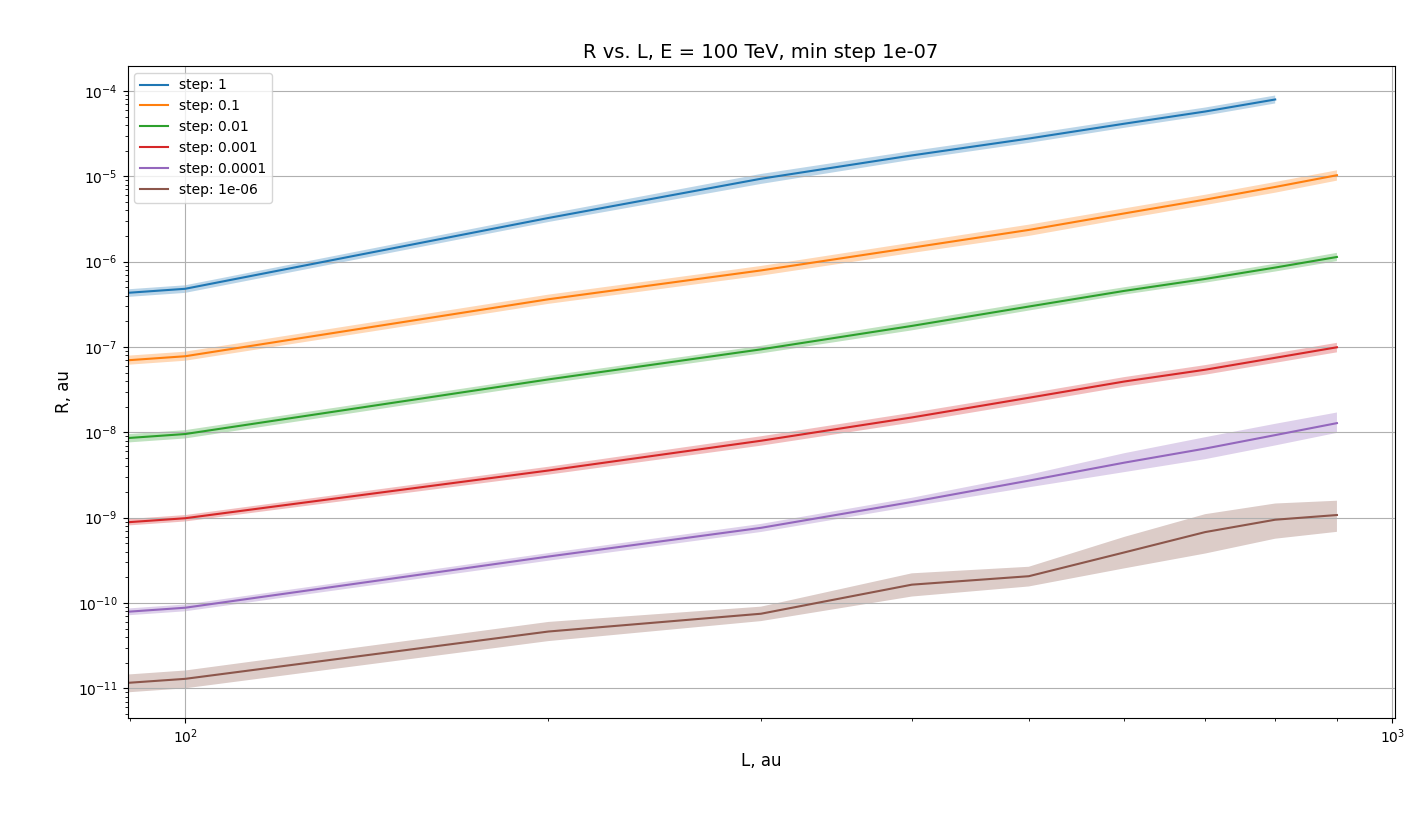}
    \caption{RMS displacement vs path length dependency.}
    \label{fig:RL_1e3au}
\end{figure}

To study numerical limits of our simulations we studied the trajectory divergence for trajectories obtained with very small step sizes. The number of trajectories used in this section is $64$, the path length is $1000$ AU, particles energies are $100$ TeV. The method of trajectory divergence calculation is the same as in Section \ref{sec:solutionStability}. The results are presented in Figure \ref{fig:RL_1e3au}. Here we can see that the difference for one selected main step ($100$ AU) between trajectories with relative step sizes $1 \cdot 10^{-7}$ and $1 \cdot 10^{-6}$ is approximately $1 \cdot 10^{-11}$. On the one hand, this difference does not equal zero while on the other hand it is nearly at the limit of our computational capabilities. So we think that the trajectory obtained with step size $1 \cdot 10^{-7}$ can represent an "ideal" trajectory for one step error estimation. 

Also we calculated the velocity trajectory divergence similar to the spatial trajectory divergence:

\begin{equation}
    \phi(l) \approx \tan(\phi(l)) = \frac{\sqrt{<(\mathbf{v}_{i}(l)-\mathbf{v}_{0}(l))^2>}}{|\mathbf{v}_{0}(l)|} = \frac{\sqrt{<(\mathbf{v}_{i}(l)-\mathbf{v}_{0}(l))^2>}}{c}
\end{equation}

\begin{figure}
    \centering
    \includegraphics[width=1\linewidth]{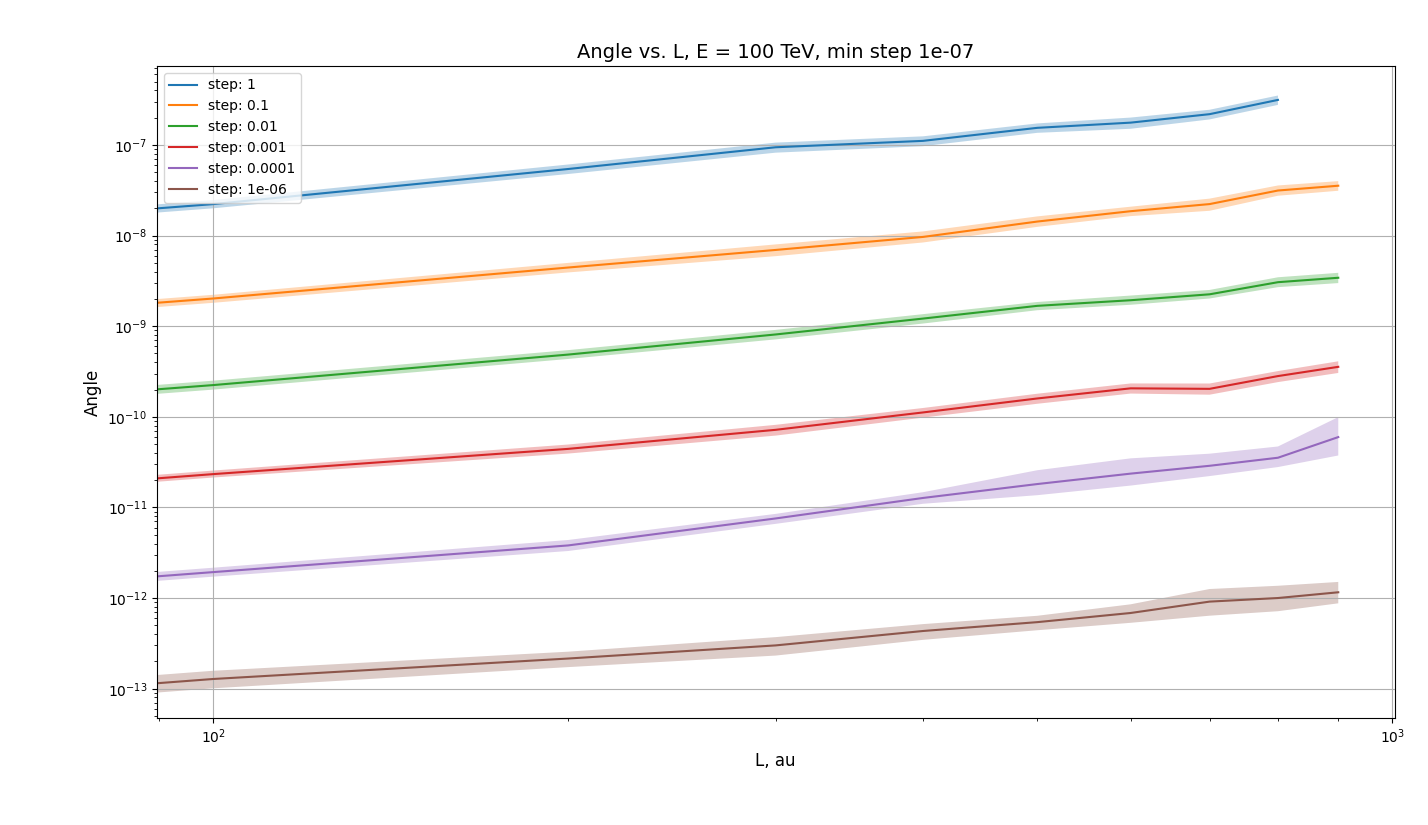}
    \caption{RMS angle between velocities vs path length dependency.}
    \label{fig:VL_1e3au}
\end{figure}

Results are presented in Figure \ref{fig:VL_1e3au}. Here we can see that the RMS angle between velocities is even closer to achieving a precision of 13-14 decimal places\footnote{Double data type limit} and this is another reason for using trajectories obtained with the step size $1 \cdot 10^{-7}$ as reference trajectories. Also, from this plot we can see that the velocity numerical error introduced by one main step is not less than $2 \cdot 10^{-8}$ radian. However, this value is a lower estimation only, because we still can not estimate how far our converging trajectories are from the true solution of the equations.

\section{Discussion and conclusion}
\label{sec:conclusion}

As shown in Section \ref{sec:solutionStability} individual trajectories obtained by numerical integration are unstable after $\sim 1$ kpc of path length. This means that there will be always some uncertainty in large-scale behavior of trajectories obtained this way, and we cannot be certain from numerical simulations alone that GCR transport has the nature of a regular diffusion. In Section \ref{sec:dStability}, the unstable behavior of the diffusion coefficient "measured" in the numerical simulations was also demonstrated. Recall that the smaller the timestep, the larger the diffusion coefficient is, and at later times (path lengths) it approaches a constant regime. This behavior is reproduced in Section \ref{sec:velocityError} in simulations that include additional velocity errors. This could mean that the diffusive behavior of GCR observed at large times is at least partially a result of numerical errors in the integration algorithm. Mathematically, these errors could introduce a cutoff in the distributions of random-walk jump sizes or accelerate pitch-angle decorrelation. The study of the jump-size distributions and intermediate pitch-angle behavior will be the focus of future work.

The observed error susceptibility of the diffusion coefficient calculations is not specific to the selected algorithm, but it is also a problem for others. Moreover, for the Boris algorithm this problem is much more severe because the causes of the trajectory divergences, and consequently the unstable behavior of the running diffusion coefficients, are local, and the algorithm has only second-order accuracy. The long-term energy stability of the Boris algorithm cannot help with the local trajectory divergences.

To sum up:

\begin{itemize}
    \item Trajectories are unstable with respect to step‑size variation.
    \item The running diffusion behavior is unstable with respect to step‑size variation.
    \item Velocity errors have a greater impact on GCR transport than spatial errors.
\end{itemize}

\section{Acknowledgements} 

The authors would like to thank Dr. Nikitin V.V. and Prof. Panov. A.D. for helpful discussions and criticism.

The research was supported by RSF (Project No.25-22-00246).

%	\newpage
%\bibliography{refs}

%\appendix

%\nocite{*}

\bibliography{apssamp.bib}% Produces the bibliography via BibTeX.

\end{document}